# *Handover Call Admission Control for Mobile Femtocells with Free-Space Optical and Macrocellular Backbone Networks*

**Mostafa Zaman Chowdhury, Nirzhar Saha, Sung Hun Chae, and Yeong Min Jang†**

Department of Electronics Engineering, Kookmin University, Korea

## Abstract

The deployment of mobile femtocellular networks can enhance the service quality for the users inside the vehicles. The deployment of mobile femtocells generates a lot of handover calls. Also, numbers of group handover scenarios are found in mobile femtocellular network deployment. The ability to seamlessly switch between the femtocells and the macrocell networks is a key concern for femtocell network deployment. However, until now there is no effective and complete handover scheme for the mobile femtocell network deployment. Also handover between the backhaul networks is a major concern for the mobile femtocellular network deployment. In this paper, we propose handover control between the access networks for the individual handover cases. Call flows for the handover between the backhaul networks of the macrocell-to-macrocell case are proposed in this paper. We also propose the link switching for the FSO based backhaul networks. The proposed resource allocation scheme ensures the negligible handover call dropping probability as well as higher bandwidth utilization.

**Key words :** Mobile femtocell, resource allocation, handover, group handover, and call dropping probability.

## 1. INTRODUCTION

Femto-access-points (FAPs) are low-power, small-size home-placed Base Stations (also known as Home NodeB or Home eNodeB) that enhance the service quality for the indoor mobile users [1-4]. Currently the mobile users inside a vehicular environment use macrocellular or the satellite networks. However, the users suffer from several difficulties e.g., low SNIR level, higher outage probability, and lower throughput due to the poor signal quality inside the vehicle. Femtocells deployment in the vehicular environment, we refer as the mobile femtocell [5] can solve these difficulties. The MS is connected to the indoor FAP instead of macrocellular or the satellite networks. A strong transceiver is installed outside the vehicle. The outside transceiver is connected to the inside FAP by a wired network and with outside macrocellular or the satellite network through a wireless link. Thus, the signal does not need to penetrate the wall of the vehicle. Therefore, the MS can receive better quality signal. Repeater (or signal booster) can also solve the poor coverage issues in vehicular environment. However, data speed of the repeater is very low compared to femtocells.

The backhauling networks are the wireless link for the mobile femtocellular networks that is the most important difference between the fixed femtocellular networks and the mobile femtocellular networks deployments. Fig. 1 shows the device-to-CN connectivity for the mobile femtocellular networks deployment. Except the backhauling networks connectivity, the other network entities are same as the fixed femtocellular networks deployment. A FAP is located inside the vehicle e.g., bus/car/train/ship or others. A transceiver is situated outside the vehicle to transmit/receive data to/from the backhaul networks (e.g., macrocellular networks, satellite networks, and etc.). The FAPs are installed inside the vehicles to make wireless connection between the users and the FAP. The FAPs and the transceiver are connected through the wired networks. The outside transceiver and the CN is connected through macrocellular networks or satellite networks or free-space optical (FSO) communication [6], [7] based backhaul networks.

†Corresponding Author: yjang@kookmin.ac.kr
Tel: +82-2-910-5068, Fax: +82-2-910-5068, Kookmin Univ.
Dept. of Electronics Engineering, Kookmin Univ., Korea

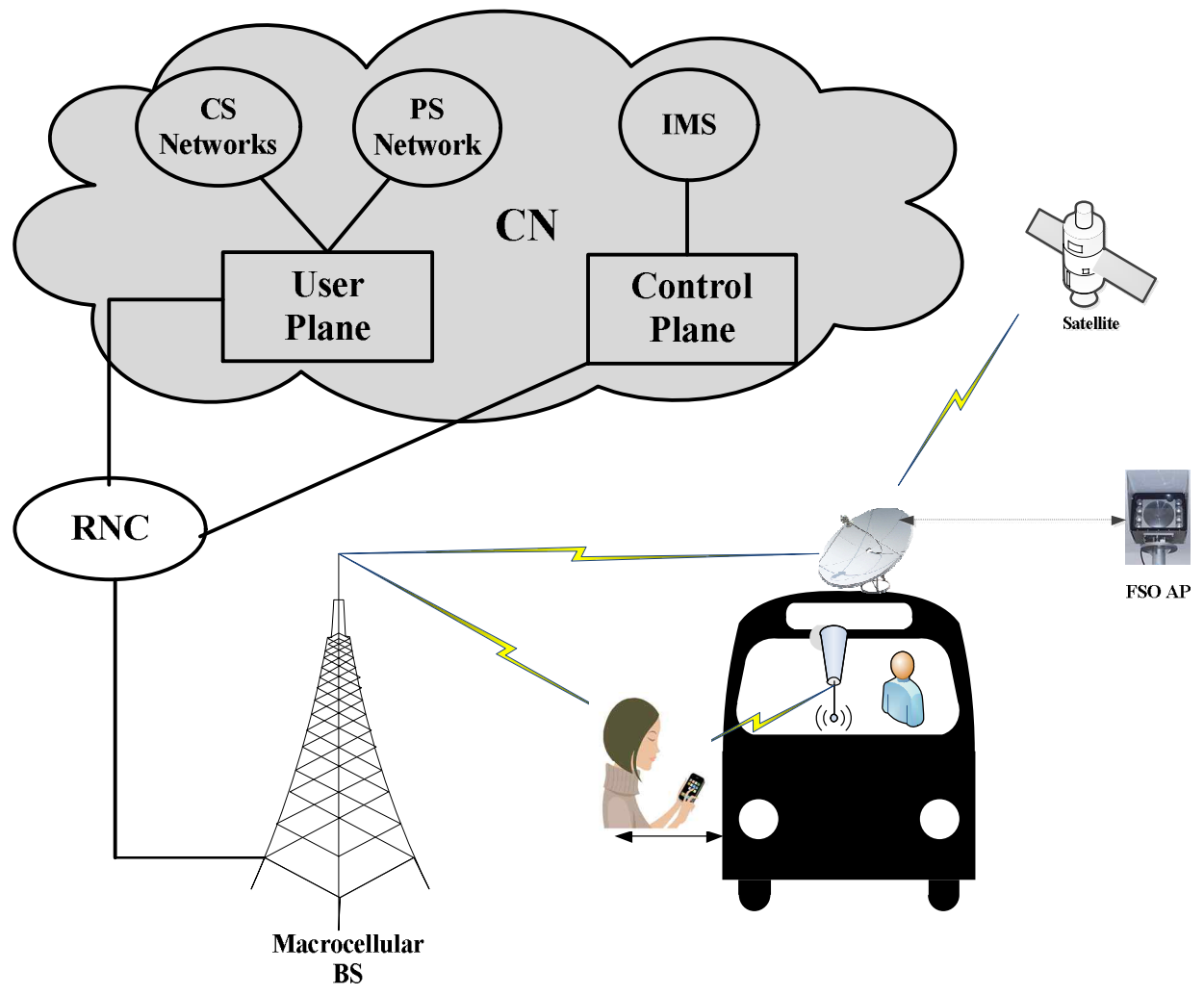

Fig. 1. Network architecture for the femtocellular network deployment in vehicular environment.

There are several issues those have to be solved to deploy femtocellular networks in vehicular environments. The backhaul network is the wireless link. This is one of the most challenging issues for the mobile femtocellular network deployment. Appropriate backhaul network selection should be considered for the design of mobile femtocellular network. The FSO communication [6], [7] based backhaul networks can provide high data rate wireless backhaul link for mobile femtocellular networks. The handover management is one of the most important issues for the mobile femtocellular network deployment. Individual handovers as well as the group handovers are happened. Sufficient amount of resources should be provided to mobile femtocellular networks during the backhaul network handovers, group handovers, and individual handovers. Therefore, we focus on the resource management for the handover in mobile femtocellular network deployment. Until now there is no effective and complete handover scheme for the mobile femtocell network deployment. Also handover between the backhaul networks is a major concern for the mobile femtocellular network deployment. In this paper, we propose handover control between the access networks for the individual handover cases. Call flows for the handover between the backhaul networks of the macrocell-to-macrocell case are proposed in this paper. We also propose the link switching for the FSO based backhaul networks. The proposed resource allocation scheme ensures the negligible handover call dropping probability as well as higher bandwidth utilization.

The rest of this paper is organized as follows. Handover controls for different scenarios are presented in Section 2. In Section 3, we provide the proposed resource allocation policy. Finally, Section 4 concludes our work.

## 2. Handover Control in Mobile Femtocellular Networks

The ability to seamlessly switch between the femtocells and the macrocell networks is a key concern for femtocell network deployment. However, until now there is no effective and complete handover scheme for the femtocell network deployment. Also handover between the backhaul networks is also a major concern for the mobile femtocellular network deployment. In this Section we propose the macrocell-to-femtocell handover and the femtocell-to macrocell handover for the individual handover for the individual handover cases based on the specifications. We also propose the call flows for the handover between the backhaul networks for the [8-12] macrocell-to macrocell case.

### *A. Femtocell-to-macrocell handover*

Fig. 2 shows the detail call flow procedures for femtocell-to-macrocell handover in mobile femtocellular network deployment. If femtocell user detects that femto signal is going down, MS sends this report to the connected FAP (steps 1, 2). The MS searches for the signals from the neighbor the macrocellular BS (step 3). Based on the pre-authentication and the received signal levels, the MS and FAP together decide for handover to the macrocellular BS (step 4). FAP starts handover (HO) procedures by sending a handover request to macrocellular BS (steps 5 and 6). The CAC and radio resource control (RRC) are performed to check whether the call can be accepted or not (steps 7, 8, and 9). Then the macrocellular BS responses for the handover request (steps 10 and 11). Steps 12, 13, and 14 are used to setup a new link between user (MS) and the macrocellular BS. The MS re-establishes a channel with the macrocellular BS and detached from the FAP, and also synchronized with the macrocellular BS (steps 15, 16, 17 and 18). MS sends a handover complete message to outside transceiver to inform that, the MS already completed handover and synchronized with the macrocellular BS (steps 19, and 20). Then the FAP deletes the old link with the MS (steps 21, and 22). Now the packets are sent to MS through the macrocellular BS.

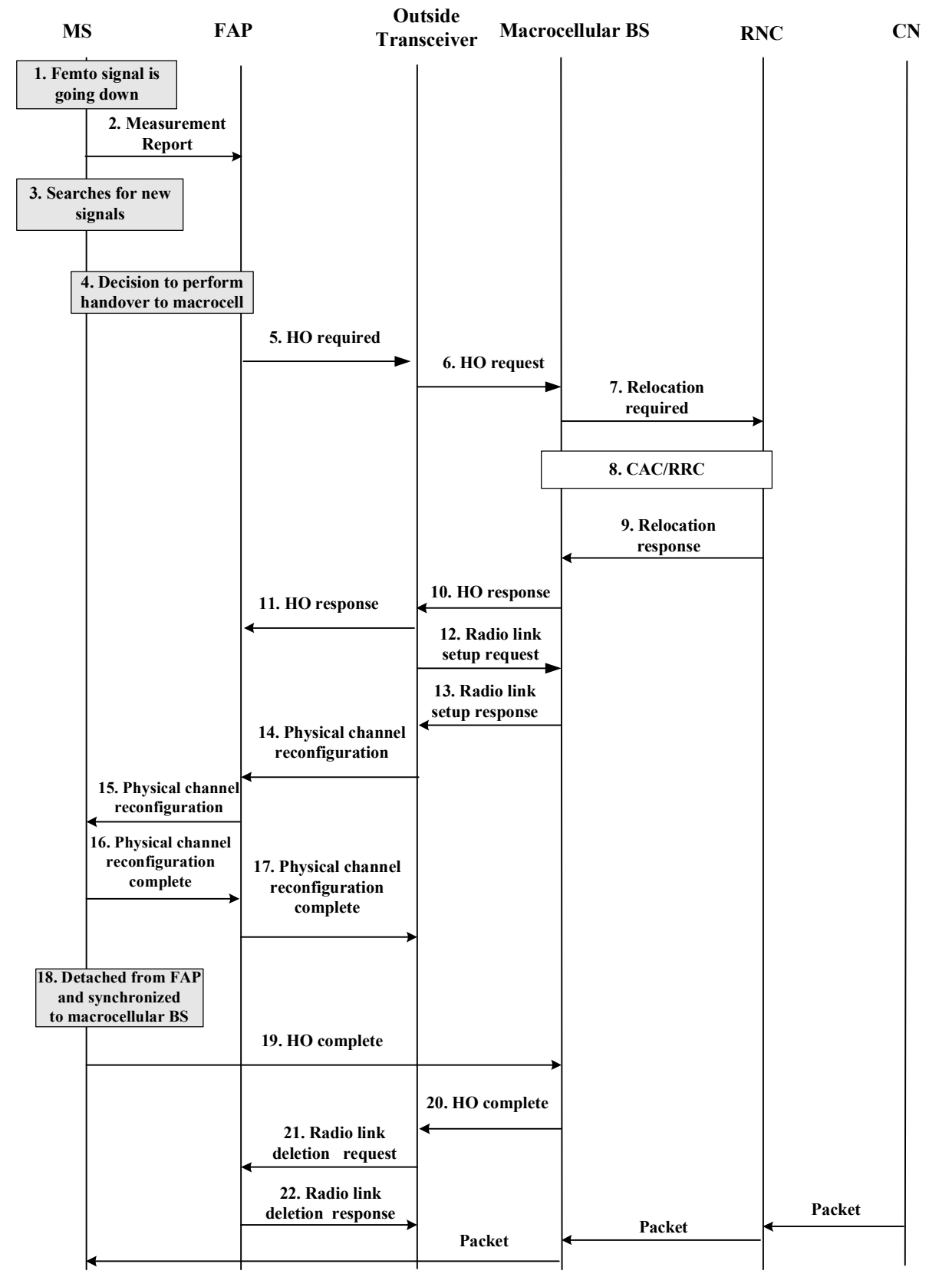

Fig. 2. Handover call flows for femtocell-to-macrocell handover in mobile femtocell environment.

### *B. Macrocell-to-femtocell handover*

Fig. 3 shows the detail call flow procedures for macrocell-to-femtocell handover in dense femtocellular network deployment. Whenever the MS in the macrocell network detects a signal from femtocell, it sends a measurement report to the connected macrocellular BS (steps 1, 2). The MS, macrocellular BS, and neighbor FAPs combine perform the SON configuration to accomplish the optimized neighbor cell list for the handover (steps 3 and 4). The MS performs the pre-authentication to all the access networks that are included in the neighbor cell list (step 5). Based on the pre-authentication and the received signal levels, the MS decides for handover to FAP (step 6). The macrocellular BS starts handover procedures by sending a handover request to the outside transceiver (step 7). The handover request is forwarded from the outside transceiver to the FAP (step 8). The FAP performs CAC, RRC, and also compares the interference levels to admit a call (steps 9). Then the FAP responses for the handover request to macrocellular BS (steps 10 and 11).

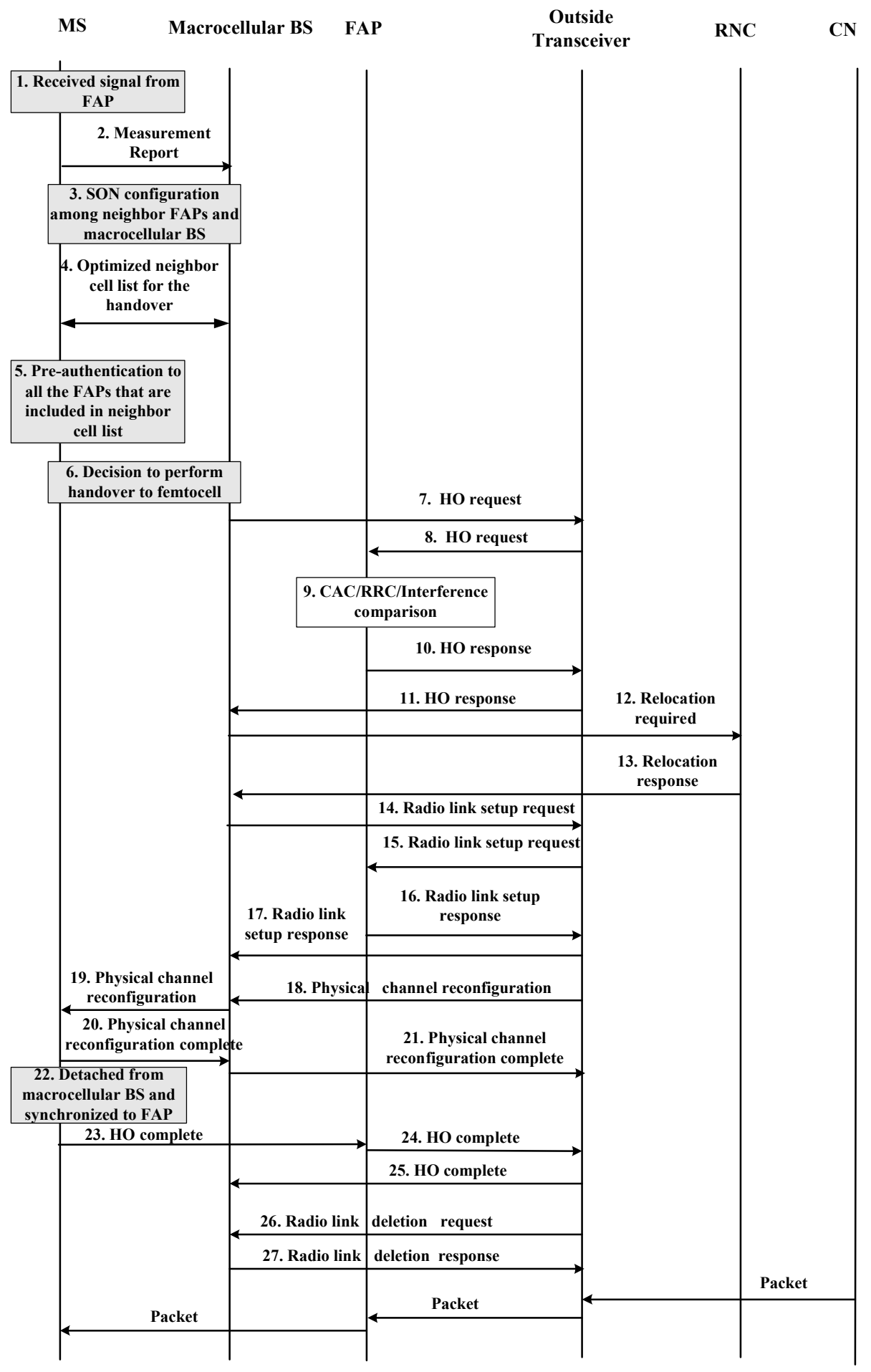

Fig. 3. Handover call flows for macrocell-to- femtocell handover in mobile femtocell environment.

A new link is established between the outside transceiver and the FAP (steps 12, 13, 14, 15, 16, and 17). Now the MS re-establishes a channel with the FAP, detached from the source macrocellular BS, and synchronized with the FAP (steps 18, 19, 20, 21, and 22). MS sends a handover complete message to outside transceiver to inform that, the MS already completed handover and synchronized with the FAP (steps 23, 24, and 25). Then the macrocellular BS deletes the old link with the outside transceiver (steps 26 and 27). Now the packets are forwarded to MS through the FAP.

### *C. Handover between backhaul networks*

The handover between the backhaul networks is an important issue for the mobile femtocellular network deployment. Various handovers can be happened, as:

- Macrocell-to-macrocell
- Satellite-to-macrocell
- Macrocell-to-satellite
- Satellite-to-satellite

- Between FSO networks and other radio networks

In this section we have shown only the macrocell-to-macrocell handover procedures. Fig. 4 shows the detail call flow procedures for macrocell-to-macrocell backhaul networks handover procedures. In this handover outside transceiver needs to select the appropriate backhaul network. If outside transceiver detects that signal is going down, outside transceiver sends this report to the connected macrocellular BS (steps 1, 2). The outside transceiver searches for the signals from the neighbor the macrocellular BSs (step 3). The outside transceiver, serving macrocellular BS, and the target macrocellular BS together perform the SON configuration to accomplish the optimized neighbor cell list for the handover (steps 4 and 5).

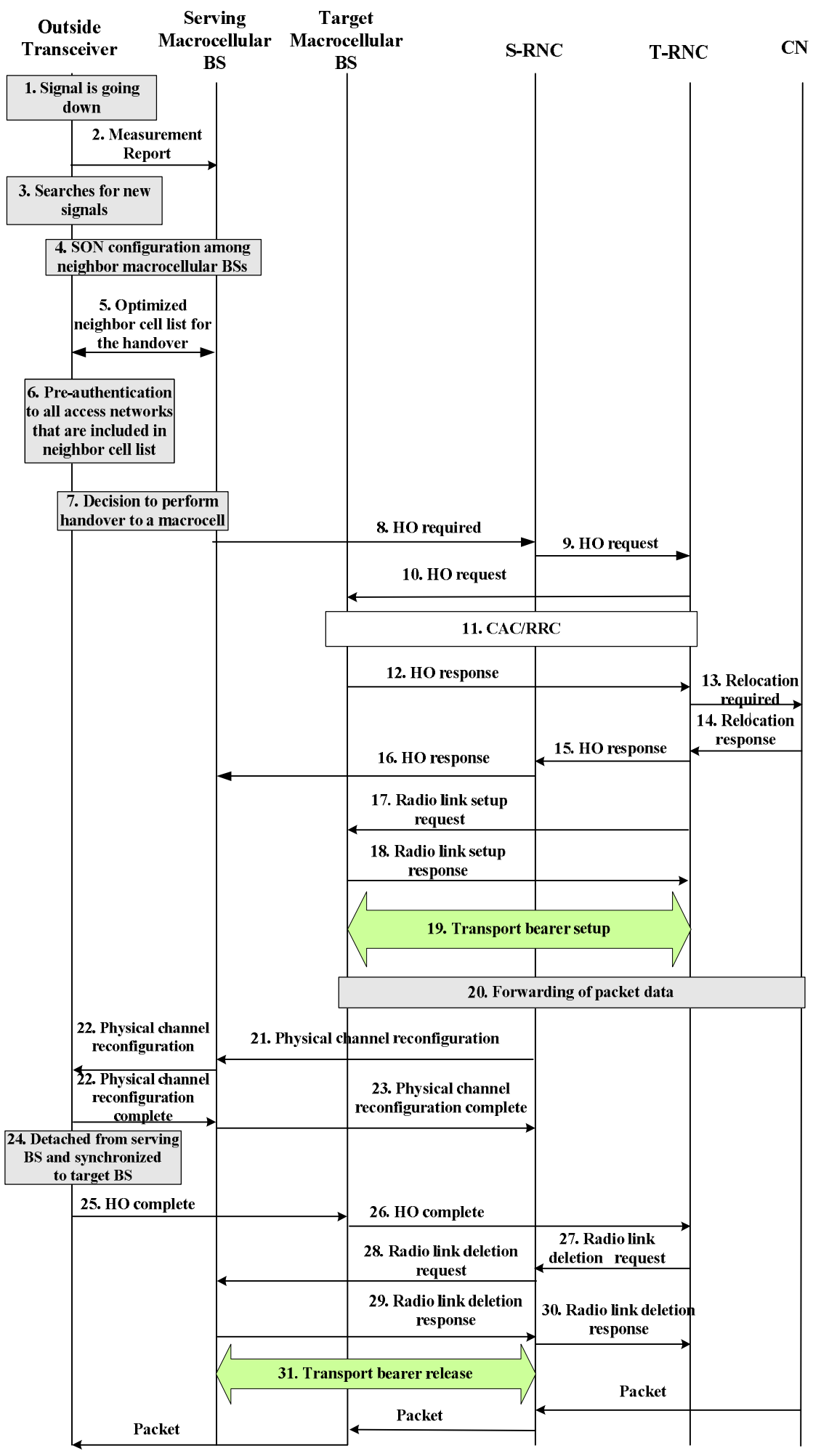

Fig. 4. Handover call flows for macrocell-to macrocell backhaul handover in mobile femtocell environment.

The outside transceiver performs the pre-authentication to all the access networks that are included in the neighbor cell list (step 6). Based on the pre-authentication and the received signal levels, the outside transceiver and serving macrocellular BS together decide for handover to the target macrocellular BS (step 7). Serving macrocellular BS starts handover procedures by sending a handover request to target serving macrocellular BS (steps 8, 9, and 10). The target macrocellular BS and target RNC perform CAC and RRC to admit the handover call (steps 11). Then the target macrocellular BS responses for the handover request to serving macrocellular BS (step 12, 13, 14, 15, and 16). A new link is established between the T-RNC and target macrocellular BS (steps 17, 18, and 19). Now the outside transceiver re-establishes a channel with the target macrocellular BS, detached from the serving macrocellular BS, and synchronized with the target macrocellular BS (steps 21, 22, 23, and 24). Outside transceiver sends a handover complete message to T-RNC to inform that, the outside transceiver already completed handover and synchronized with the target macrocellular BS (steps 25 and 26). Then the serving macrocellular BS deletes the old link with S-RNC (steps 27, 28, 29, 30, and 31). Now the packets are forwarded to outside transceiver through the target macrocellular BS.

## *D. Link switching for the FSO backhaul networks*

The FSO communication networks can be installed at highway road or at train line or at subway line. This backhauling connection provides very high data rate for the mobile femtocellular networks. The link switching is happened when the backhaul network is changed from one FSO AP to another FSO AP. Link switching is needed due to mobility of the vehicle. Fig. 5 shows the signal flow processing between the optical transceiver at vehicle and the target FSO AP during link switching. The optical channel gain that is related to transmitted and received powers can be expressed as:

$$P_r = H(0)P_t \tag{1}$$

where $P_t$ is the transmitted optical power, $P_r$ is the received optical power, and $H(0)$ is the channel DC gain.

Considering the LOS link, the channel DC gain is [13]:

$$H_{LOS} = \begin{cases} \dfrac{(\tau+1)A}{2\pi D^2}\cos^{\tau}(\varphi)T_S(\psi)g(\psi)\cos(\psi), & 0 \le \psi \le \psi_c \\ 0, & elsewhere \end{cases} \tag{2}$$

where $\tau$ is the order of Lambertian emission, $A$ is the photo-detector area, $D$ is the distance between the transmitter and receiver, $\varphi$ is the angle of irradiance, $\psi$ is the angle of incidence, $T_s(\psi)$ is the signal transmission coefficient of an optical filter, $g(\psi)$ is the gain of an optical concentrator, and $\psi_c$ is the receiver field of view (FOV).

The order of Lambertian emission $\tau$ can be found from the equation,

$$\tau = -\frac{ln\,2}{ln(cos\,\phi_{1/2})} \tag{3}$$

where $\phi_{1/2}$ is the transmitter half power angle. The gain can be determined from the following expression [4]:

$$g(\psi) = \begin{cases} \frac{v^2}{\sin^2(\psi_c)}, & 0 \le \psi \le \psi_c \\ 0, & elsewhere \end{cases} \tag{4}$$

where $v$ denotes the internal refractive index of the optical concentrator.

Fig. 6 shows the delay performance measurement for link switching. We chose OMNET, one of believable network analyzing tools for our simulation. It seems that the link switching delay is 136 ms.

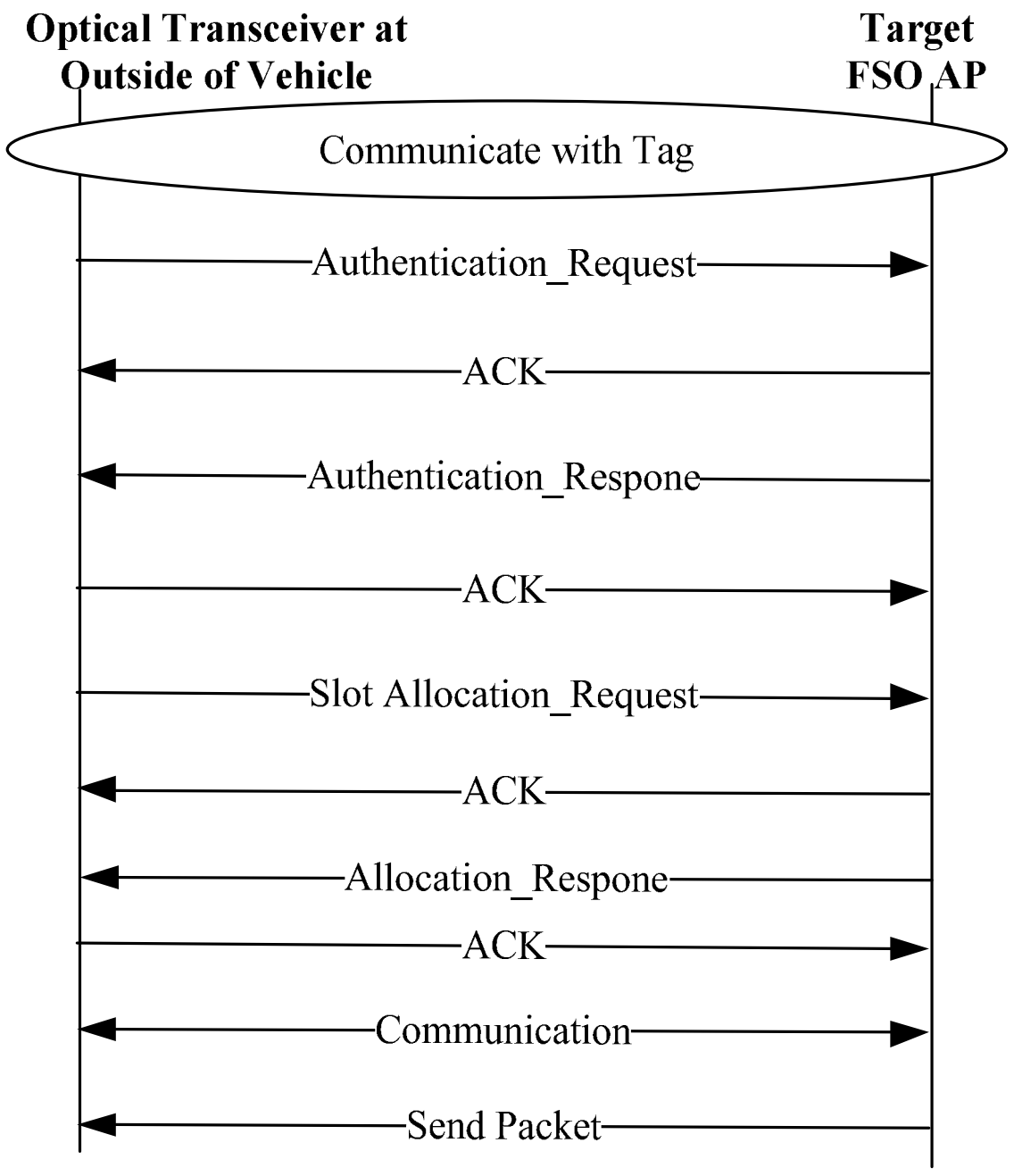

Fig. 5. Signal flow processing.

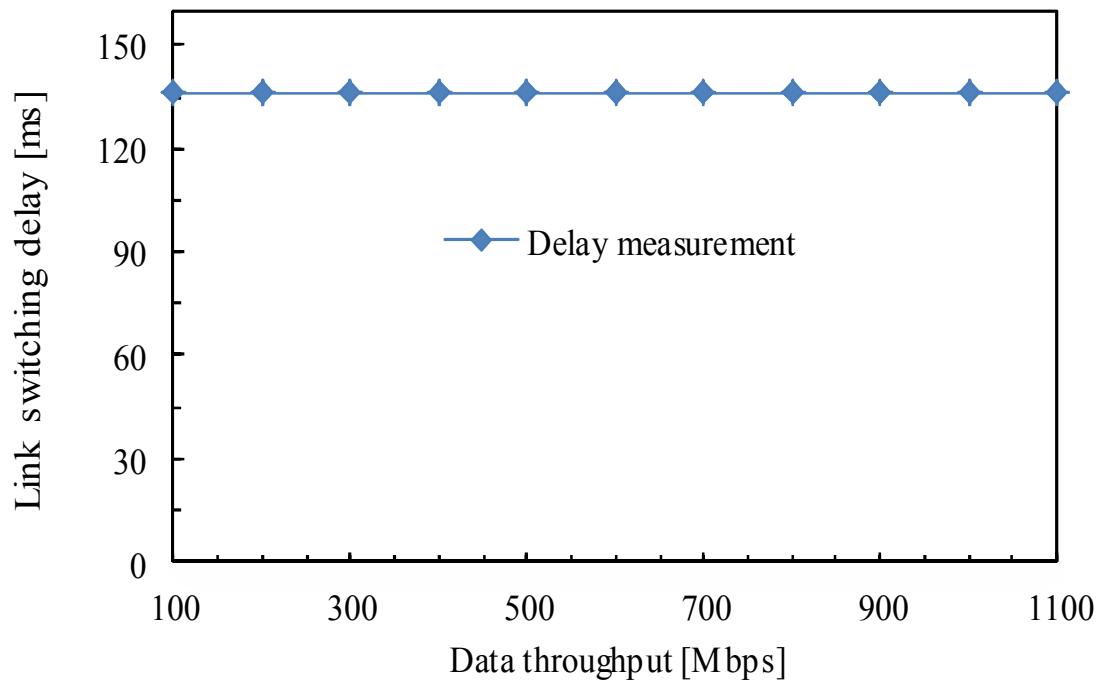

Fig. 6. Link switching delay measurement.

## 3. RESOURCE ALLOCATION IN MACROCELLULAR NETWORKS

This section represents an example of radio resource management scheme to support handover calls considering the macrocellular networks as the backhauling networks. Radio resource allocation is a very important issue in macrocellular networks for handover in mobile femtocellular network. The femtocell-to-macrocell handover rate is increased because of mobile femtocellular network deployment. Generally a group of users connecting to a mobile femtocell move away from one macrocell coverage to another macrocell coverage. Hence it is necessary for macrocell system to accommodate the movement of mobile femtocells. Otherwise all the users will lose their connections with the femtocell. Efficient radio resource management scheme is required in order to support these handover calls. The proposed radio resource allocation scheme in macrocellular networks for handovers in mobile femtocellular network deployment is shown in fig. 7. According to our proposed scheme, $C_{vacant}$, represents the vacant bandwidth due to macrocell-to-femtocell handover and due to the movement of mobile femtocellular networks from one macrocell to another macrocell during time T. $C_{occupied}$ and $C_{releasable}$ are, respectively, the occupied bandwidth by the existing calls, and the amount of releasable bandwidth from the existing non-real-time calls. Therefore, the priority of the handover calls in the macrocellular networks is given by few reserved bandwidth and the QoS adaptation.

The scheme we proposed, allows the adaptation of QoS adaptive calls. The system can release a fixed amount of bandwidth from the running QoS adaptive calls to accommodate the handover calls. In our scheme, we mentioned this recoverable bandwidth as $C_{releasable}$. The QoS adaptability [14] of some multimedia traffic calls allows the regaining of system bandwidth to support as many handover calls as possible. The releasable amount of bandwidth from QoS adaptive call is calculated as:

$$\beta_{releasable(i)} = \xi_i \beta_{r,i} \qquad (5)$$

$$C_{releasable} = \sum_{i=1}^{M} N_i \xi_i \beta_{r,i} \qquad (6)$$

where $\xi_i$ is designated as the maximum portion of bandwidth that can be degraded for a background traffic call of *i-th* class to accept a handover call in the system. $\beta_{r,i}$ is the requested bandwidth for a background traffic call of class *i*. *M* is the total number of existing QoS adaptive traffic class. $N_i$ is the total number of existing calls of traffic class *i*.

When any user moves between mobile femtocell and macrocell, the freed bandwidth due to the movement of the user from the macrocell-to-femtocell handover is reserved for the time *T* to accommodate handover calls. Furthermore, when the mobile femtocell moves from one macrocell to another macrocell, a mobile femtocell moves from one macrocell to another macrocell coverage area, the freed bandwidth from the previous macrocellular network will be reserved for the threshold time *T*.

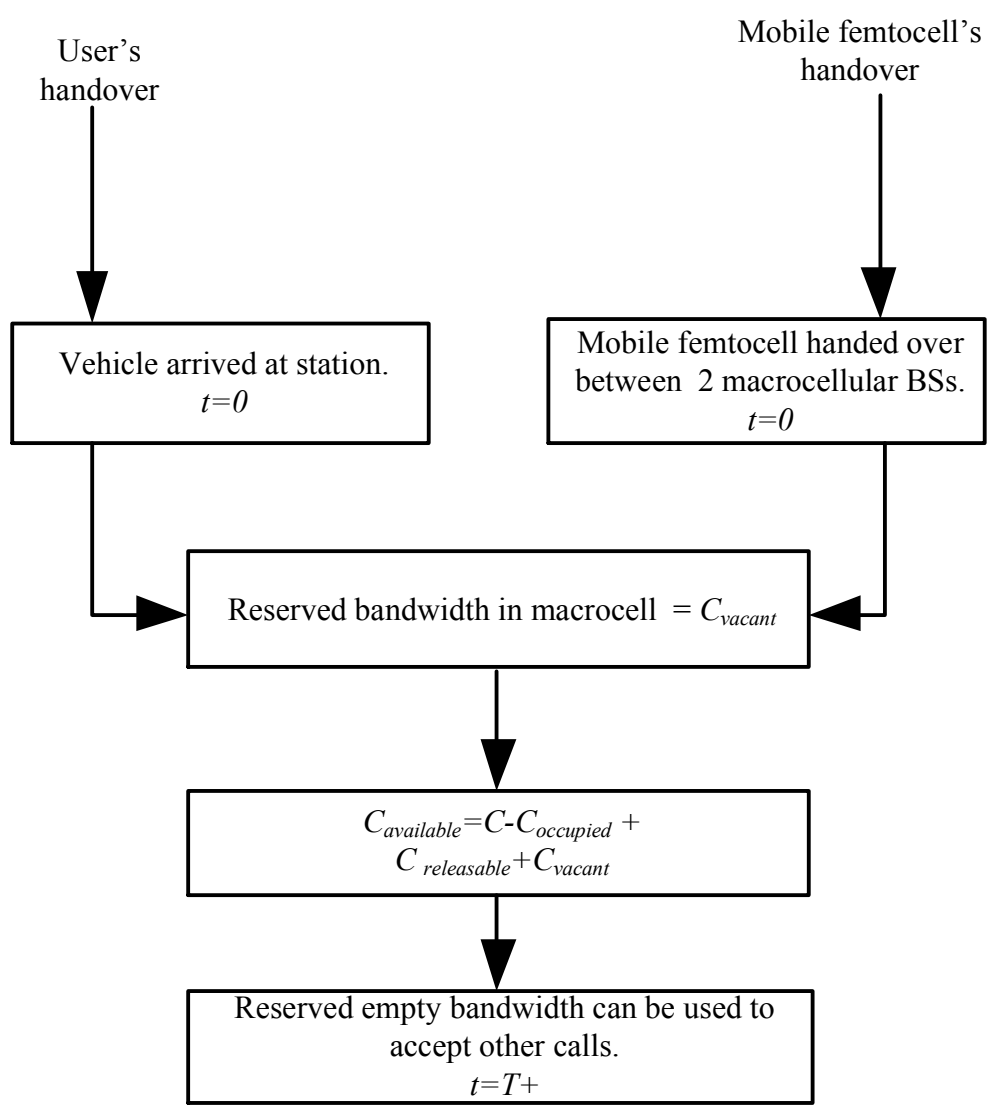

Fig. 7. Radio resource allocation in macrocellular networks to support handover calls in mobile femtocellular network deployment.

Thus the, dynamically variable $C_{vacant}$ amount of bandwidth is reserved to accommodate the handover calls. The prime feature of this bandwidth reservation scheme is that, the reserved amount of bandwidth is increased if the traffic due to the mobile femtocells is increased.

The propagation model for a macrocell case can be expressed as:

$$\begin{aligned} L = 36.55 + 26.16\log_{10} f_c - 3.82\log_{10} h_b - a(h_m) \\ + \left[44.9 - 6.55\log_{10} h_b\right]\log_{10} d + L_{sh} + L_{pen}\,[dB] \end{aligned} \qquad (7)$$

$$a(h_m) = 1.1[\log_{10} f_{c,m} - 0.7]h_m - (1.56\log_{10} f_{c,m} - 0.8) \qquad (8)$$

where $f_{c,m}$ is the center frequency in MHz of the macrocell, $h_b$ is the height if the macrocellular BS in meter, $h_m$ is the height of the MS in meter, *d* is the distance between the macrocellular BS and the MS in kilometer, $L_{sh}$ is the shadowing standard deviation, and $L_{pen}$ is the penetration loss.

The propagation model for a femtocell case can be expressed as:

$$L_{fem} = 20\log_{10} f_{c,f} + N\log_{10} d_1 + 4n^2 - 28 \;\; [dB] \qquad (9)$$

where $f_{c,f}$ is the center frequency in MHz of the femtocell, *n* is the number of walls between the MS and the FAP, and $d_l$ is the distance between the FAP and the MS in meter.

We consider the macrocellular networks as the backhauling networks in our analysis. The values of the parameters that we used in our simulation are given in Table I.

Table I: Summary of the parameter values used in analysis

| Parameter | Value |
|---|---|
| Bandwidth capacity of macrocellular networks | 6 Mbps |
| Maximum portion of bandwidth that can be degraded for a QoS adaptive traffic call of *i-th* class ($\xi_i$) | 0.5 |
| Ratio of QoS adaptive traffic and non-QoS adaptive traffic | 1 : 1 |
| Macrocell-to-macrocell handover call arrival rate : mobile femtocell-to-macrocell handover call arrival rate | 1 : 1 |
| Average call duration time considering all calls (exponentially distributed) | 120 sec |
| Average cell dwell time for the macrocell users (exponentially distributed) | 540 sec |
| Threshold time for bandwidth reservation (*T*) | 10 sec |

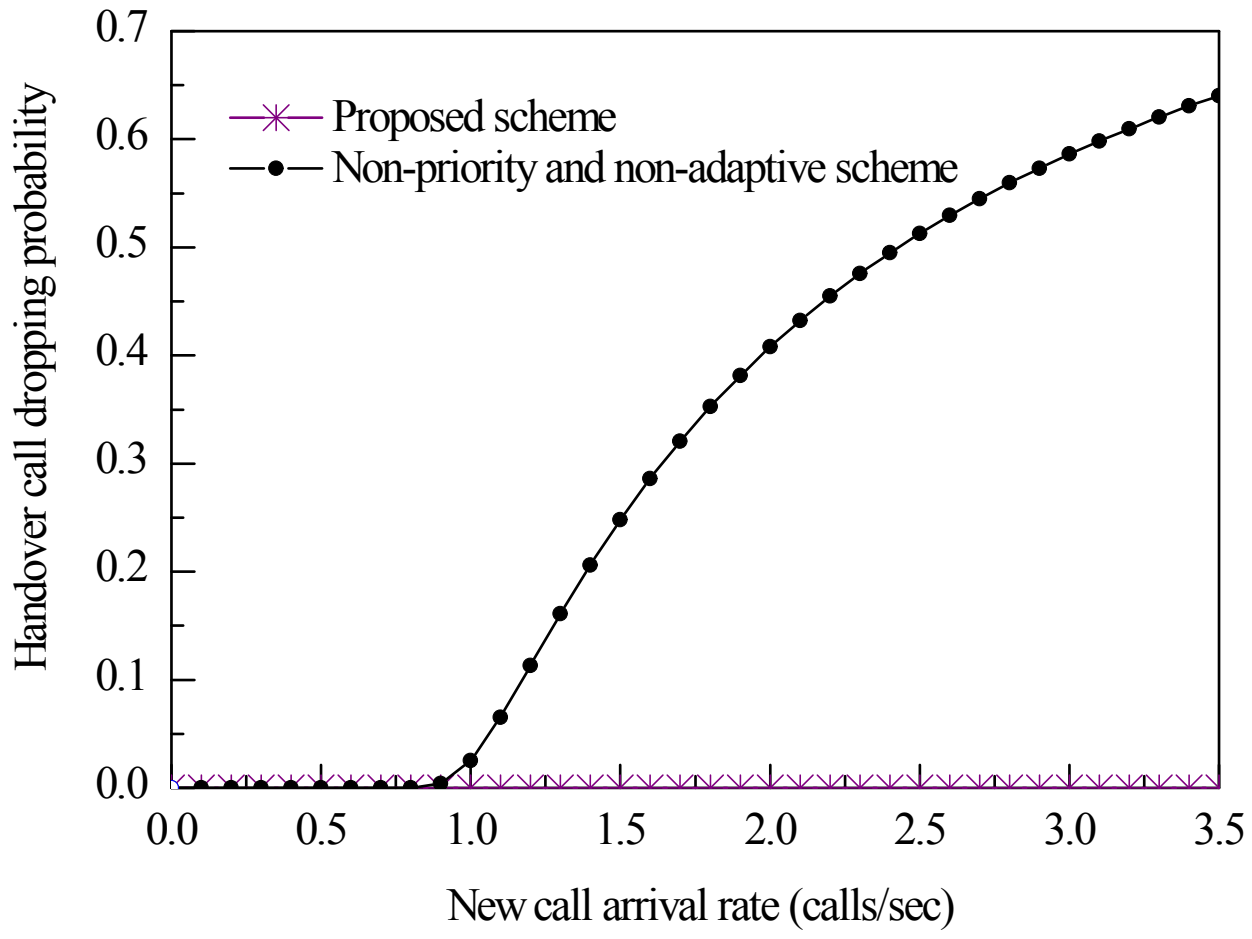

Fig. 8. Comparison of handover call dropping probability.

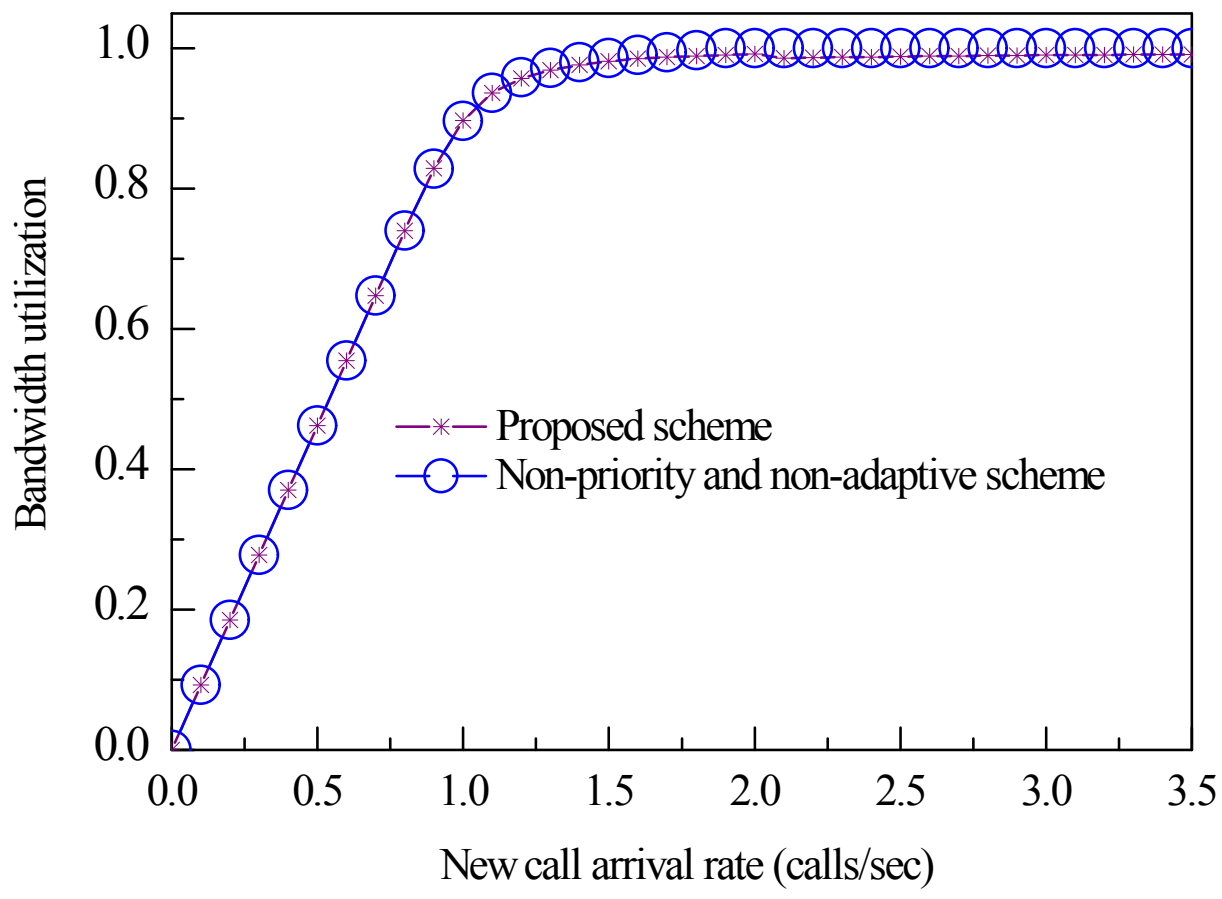

Fig. 9. Comparison of bandwidth utilization.

Fig. 8 shows that the proposed scheme provides negligible handover call dropping probability even for very high traffic condition. The other scheme cannot provide lower handover call dropping probability due to the non-priority of handover calls. Fig. 9 shows the bandwidth utilization comparison. Even though our scheme contains bandwidth reservation and the priority of the handover calls, the proposed scheme does not significantly reduce the bandwidth utilization.

## 4. Conclusion

The mobile femtocells will be the new paradigm of the femtocellular network deployment. However, the deployment of mobile femtocells will create a large number of handover calls within the macrocellular networks. Therefore, the handover management is one of the key issues for the successful mobile femtocell network deployment. We study a complete lesson for the challenging problems of handover management in an mobile femtocell network. We proposed several solutions for the handover issues in mobile femtocellular network deployment. The performance analysis shows the advantages of our proposed scheme.

## Acknowledgment

This work was supported by the IT R&D program of MKE/KEIT [10035362, Development of Home Network Technology based on LED-ID].

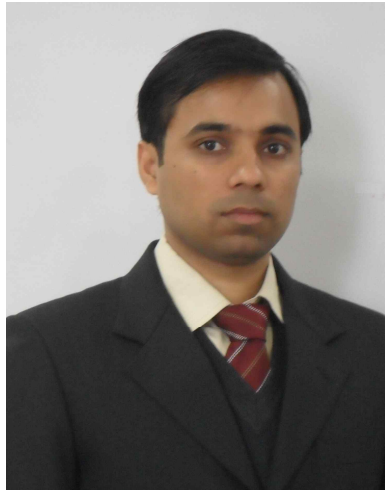

**Mostafa Zaman Chowdhury** received his B.Sc. in Electrical and Electronic Engineering from Khulna University of Engineering and Technology (KUET), Bangladesh in 2002. In 2003, he joined the Electrical and Electronic Engineering department of KUET as a faculty member. He received his M.Sc. in Electronics Engineering from Kookmin University, Korea in 2008. Currently he is working towards his Ph.D. degree in the department of Electronics Engineering at the Kookmin University, Korea. His research interests include convergence networks, QoS provisioning, mobility management, femtocell networks, and VLC networks.

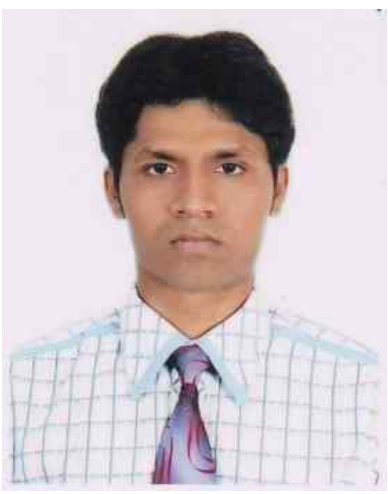

**Nirzhar Saha** received his B.Sc. in Electrical and Electronic Engineering from Khulna University of Engineering and Technology (KUET), Bangladesh in 2011. In 2012, he joined the wireless network and communication lab as a research student in the department of Electronics Engineering at Kookmin University, Korea. His research interests include visible light communication networks, OFDM and multi-carrier systems, cognitive radio networks.

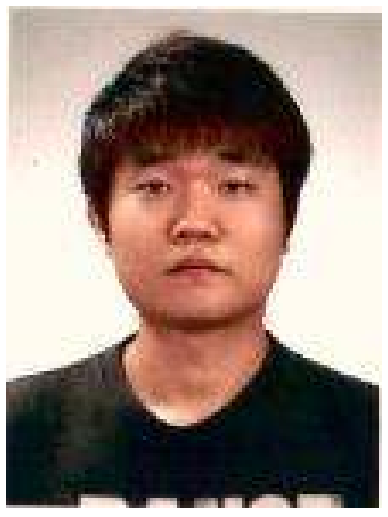

**Sung Hun Chae** received his B.Sc. in Electrical Engineering from Kookmin Kookmin University, Korea in 2012. Currently he is working towards his M.Sc. degree in the department of Electronics Engineering at the Kookmin University, Korea. His research interests include QoS provisioning, mobility management and femtocell networks.

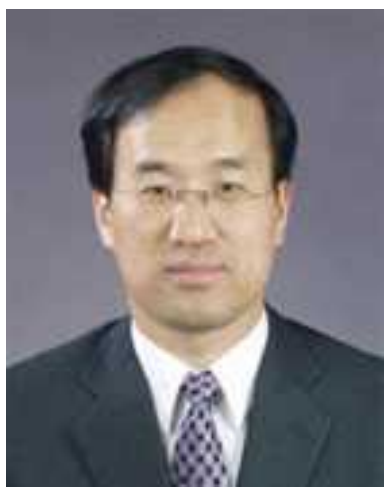

**Yeong Min Jang** received the B.E. and M.E. degree in Electronics Engineering from Kyungpook National University, Korea, in 1985 and 1987, respectively. He received the doctoral degree in Computer Science from the University of Massachusetts, USA, in 1999. He worked for ETRI between 1987 and 2000. Since September 2002, he is with the School of Electrical Engineering, Kookmin University, Seoul, Korea. His research interests include IMT-advanced, radio resource management, femtocell networks, Multi-screen convergence networks, and WPANs.